\documentclass[a4paper,11pt]{article}
\usepackage{pos}

\usepackage[utf8]{inputenc}
\usepackage[backend=biber,style=chem-angew,maxbibnames=1,doi=true,isbn=false,url=false,eprint=false,sorting=none]{biblatex}
\AtEveryBibitem{\clearlist{publisher}}
\AtEveryBibitem{\clearlist{pages}}
\AtEveryBibitem{\clearfield{pages}}
\AtEveryBibitem{\clearfield{publisher}}
\AtEveryBibitem{
    \clearfield{urlyear}
    \clearfield{urlmonth}
}

\usepackage{caption}
\usepackage[T1]{fontenc}    
\usepackage{hyperref} 
\usepackage{url}            
\usepackage{booktabs}       
\usepackage{amsfonts}       
\usepackage{nicefrac}       
\usepackage{microtype}      
\usepackage{lipsum}
\usepackage{graphicx}
\graphicspath{ {./figures/} }
\usepackage{wrapfig}
\usepackage{xspace}
\usepackage{amsmath}
\usepackage{amstext}

\usepackage{todonotes}
\newcommand{\txs}{TXS~0506+056\xspace}
\newcommand{\ngc}{NGC~1068\xspace}
\newcommand{\ao}{AO~0235+164\xspace}
\newcommand{\gwgrb}{GW170817-GRB\xspace}
\newcommand{\pks}{PKS~1502+106\xspace}

\newcommand{\plenum}{PLEnuM\xspace} 

\usepackage{siunitx}

\DeclareSIUnit{\EeV}{EeV}
\DeclareSIUnit{\PeV}{PeV}
\DeclareSIUnit{\TeV}{TeV}
\DeclareSIUnit{\Mpc}{Mpc}
\DeclareSIUnit{\Gpc}{Gpc}
\DeclareSIUnit{\erg}{erg}
\DeclareSIUnit{\year}{yr}

\title{Prospects for the detection of transient neutrino sources with PLEnuM}
\author*[a]{Lisa Johanna Schumacher}
\author[b]{Foteini Oikonomou}

\affiliation[a]{Erlangen Centre of Astroparticle Physics, FAU Erlangen-Nürnberg, Germany}

\affiliation[b]{Norwegian University of Science and Technology, Trondheim, Norway}

\emailAdd{lisa.j.schumacher@fau.de}
\emailAdd{foteini.oikonomou@ntnu.no}

\abstract{
The discovery of high-energy astrophysical neutrinos in the TeV-PeV range by IceCube marked the start of neutrino astronomy, and the search for their sources continues.
Two promising source candidates have been identified by IceCube: \ngc in the 1 TeV-10 TeV range and \txs in the 0.1-1 PeV range.
Both sources have gamma-ray counterparts, but additional time information of both neutrinos and gamma rays were essential for the identification of \txs.
The Planetary Neutrino Monitoring (\plenum) concept is an approach for combining the exposures of all current and future neutrino observatories - such as KM3NeT, Baikal-GVD, P-ONE in the Northern Hemisphere, and IceCube-Gen2 in the Southern Hemisphere.
Using this \plenum approach, we estimate how the detection capability for transient sources candidates like blazars and GRBs improves once the future neutrino observatories come online.
In addition, we present how the combined, instantaneous field of view of \plenum improves the real-time detection rate of rare, very-high-energy neutrinos across the entire sky.
}

\ConferenceLogo{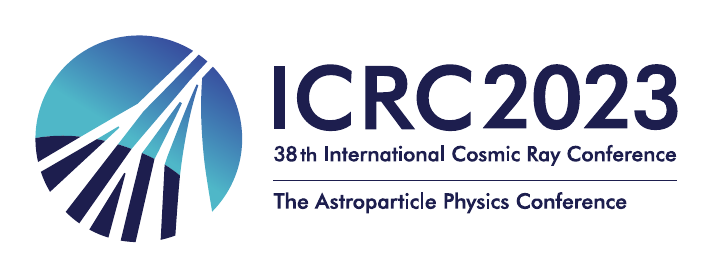}

\FullConference{%
38th International Cosmic Ray Conference (ICRC2023)\\
  26 July - 3 August, 2023\\
  Nagoya, Japan}

\begin{document}

\maketitle

\section{Introduction}
Over the coming decades, numerous neutrino telescopes around the world will grow and become fully operational:
KM3NeT~\cite{km3netIntent2016}, P-ONE~\cite{agostini_pacific_2020}, and Baikal-GVD~\cite{baikalGVD2019} in the Northern Hemisphere, and IceCube-Gen2~\cite{Aartsen_2021:gen2} at the South Pole as an extension to IceCube.
These detectors together will greatly increase the detection rate of high-energy, astrophysical neutrinos compared to what IceCube achieved in its more than \SI{10}{years} of operation time.
We introduce here the Planetary Neutrino Monitoring (\plenum), which is a concept for combining the high-energy neutrino observations made by current and future neutrino telescopes in order to answer the open questions in the field of neutrino astronomy. 
In this proceeding, we focus on the capabilities of \plenum to observe transient sources that produce neutrinos at the highest energies.

All calculations presented here are based on the combined effective areas of the neutrino telescopes at different locations on the globe.
For our calculations, we currently use only the effective area of the IC86-II data of IceCube from the data release of muon neutrino events~\cite{ICdataRelease2021arxiv}.
This data set is ideal for point-source searches due to the sub-degree angular resolution of the muon neutrino events above \SI{10}{TeV}.
For the sake of simplicity, the other detectors in the Northern Hemisphere (KM3NeT, P-ONE, Baikal-GVD) are modeled by rotating the effective area of IceCube to these different locations on the globe.
Future work will include the detectors' actual effective areas and resolutions.
We investigate here the capabilities of two different configurations of \plenum:
\plenum-1 consists of IceCube together with all three Northern-Hemisphere telescopes.
For \plenum-2, we model the contribution of IceCube-Gen2 at the South Pole with a detector that has a 7.5 times larger effective area than IceCube, while keeping the resolution functions the same as for IceCube.
The effective areas are shown in~\cite{schumacher_plenum_2022}, Fig.~1. 
The code of this work is open source and available on GitHub at \href{https://github.com/PLEnuM-group/Plenum}{\path{https://github.com/PLEnuM-group/Plenum}}.

\section{Method}
\subsection{Flare models}
\label{sec:models}
We study the detectability of bright neutrino flares produced by powerful blazars and neutrino emission coincident with short gamma-ray bursts (GRBs). 
Blazar flares are promising periods of neutrino emission, as often blazars release a significant fraction of their total energy output during such epochs~\cite{Yoshida:2022wac}. 
Here, we consider neutrino emission from two of the brightest flat spectrum radio quasars \ao, and \pks. 
The latter has been associated with a high-energy muon neutrino detected with IceCube~\cite{rodrigues_multiwavelength_2021,Oikonomou:2021akf}. 
Both of these sources have produced strong gamma-ray flares during the last decade, and were among a handful of the brightest gamma-ray emitting objects in the sky during the flaring epochs. 
We also simulate neutrino emission similar to that detected in the direction of \txs in 2014-15~\cite{icecubePriorTXS2018}. 
We base our calculation on the broad-line region model of~\cite{rodrigues_multiwavelength_2021}, which is consistent with the estimate of the broad-line-region luminosity of~\cite{Padovani:2019_masquerading}.

Short GRBs (sGRBs) are the only astrophysical phenomena that have been associated with gravitational waves and electromagnetic counterpart emission to date~\cite{the_ligo_scientific_collaboration_gw170817_2017}. 
sGRBs are also promising neutrino sources~\cite{fang_high-energy_2017,Kimura:2017kan} due to the fact that the merger environment is rich in gas,
In addition, there are intense photon fields associated with the GRB, which can produce particles with extremely high energies. 
Nevertheless, IceCube and ANTARES have placed strong upper limits on neutrino production in sGRBs which strongly constrain the most optimistic neutrino production models~\cite{albert_search_2017,ANTARES:2020vzs,abbasi_searches_2022}. 
Here, we study the detectability of neutrinos from sGRBs similar to \gwgrb with \plenum, based on the post-merger magnetar scenario of~\cite{fang_high-energy_2017} in which neutrinos are produced in interactions of accelerated particles with thermal photons in the pulsar wind nebula following the merger. 
To do so, we scale the model in~\cite{fang_high-energy_2017} to the distance of \gwgrb of \SI{40}{Mpc}, as has been done in~\cite{albert_search_2017}. 

The subset of sources we investigate in the following is selected to cover a range of transients likely to produce high-energy neutrinos:
\begin{itemize}
    \item \textbf{\ao} with a flare duration of \SI{84}{days} has the highest neutrino flux prediction in the blazar selection of~\cite{oikonomou_high-energy_2019}; the expected number of detected neutrino events is $n_\nu \approx 0.07$.
    \item \textbf{\gwgrb} has a neutrino prediction for the 14-day period of $n_\nu \approx 0.04$ and a spectrum that extends to the highest energies~\cite{fang_high-energy_2017}
    \item \textbf{\pks} has the highest neutrino-flux prediction in~\cite{rodrigues_multiwavelength_2021}, yielding $n_\nu \approx 3$, based on the lepto-hadronic model of the hard-spectrum state. As in~\cite{rodrigues_multiwavelength_2021}, we assume a total flaring duration of \SI{3.7}{years} within the observation time of \SI{10}{years}.
    \item \textbf{\txs} had a flare in 2014-2015 with a duration of \SI{158}{days} that has been detected by IceCube~\cite{icecubePriorTXS2018}. The flux model we use from~\cite{rodrigues_leptohadronic_2019}, as shown in Fig. 3, model a), yields $n_\nu \approx 6$ neutrinos.
\end{itemize}
All event numbers are calculated using IceCube's effective area at the source declination and the stated flare duration.
The energy spectra of the sources are here treated as templates, which are placed at different declinations to
model similar sources at different coordinates.

\subsection{High-energy neutrino alerts}
\label{sec:alerts}

\begin{figure}[h!]
    \centering
    \includegraphics[width=\textwidth]{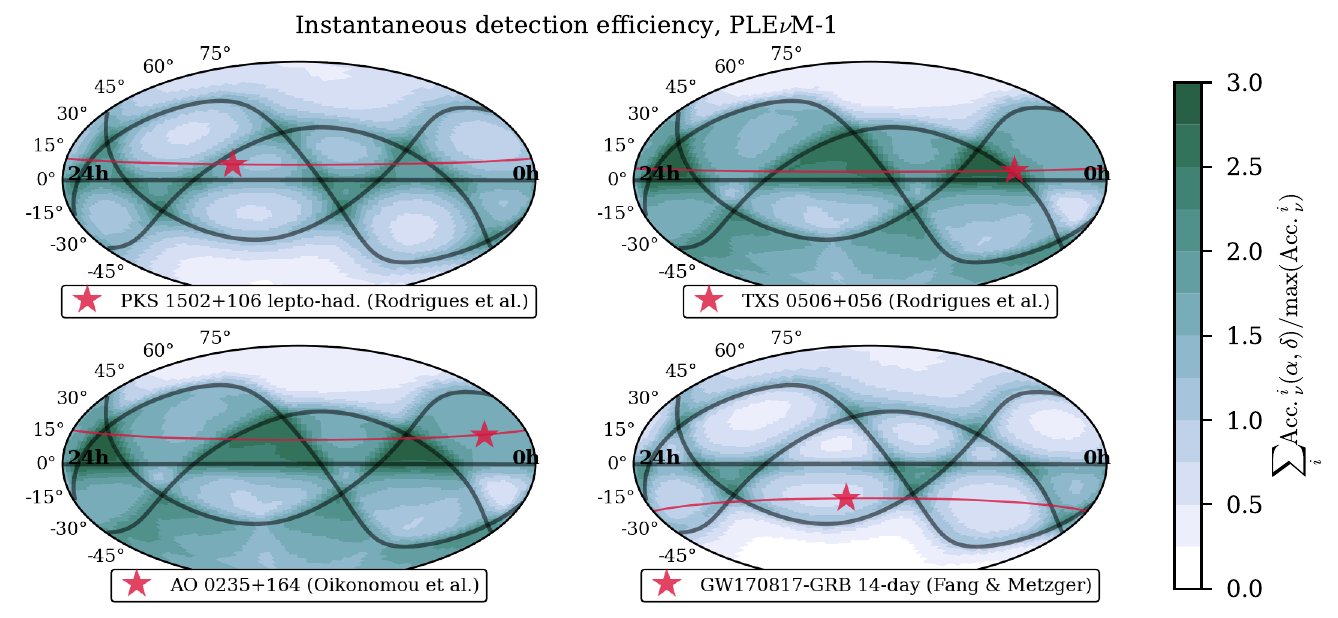}
    \caption{Instantaneous acceptance for high-energy neutrinos with energy $E_{\rm reco} > \SI{100}{TeV}$ with \plenum-1.
    The acceptance in the different panels is calculated using the energy spectrum of the respective source model, but located at different locations on the sky.
    The red stars and horizontal lines mark the actual source location and the respective declination.
    The black lines mark the local horizons of each detector at one moment in time.
    The color bar is normalized such that the maximum acceptance of one detector is 1. }
    \label{fig:inst_acceptance}
\end{figure}

One way to search for sources is to look for single, high-energy neutrinos and identify a spatial and temporal coincidence with a transient source.
This could for example be a flaring blazar like \txs~\cite{icecubeMultiMessengerTXS2018}, or a one-time event like a GRB~\cite{albert_search_2017,abbasi_limits_2023}.
While IceCube has specific event selections for these "alert" events, we approximate the distribution of arrival directions of muon neutrinos with the data release we use also for our other studies presented here. 
We model an alert selection by requiring the reconstructed energy of the secondary muon to be $E_{\rm reco} > \SI{100}{TeV}$.
For the sake of simplicity, we do not apply a cut on the zenith, neither a zenith-dependent cut on the energy as IceCube does.
This is an interesting case to study with \plenum-1, as the four detectors together can observe a larger region of the sky with a larger instantaneous signal acceptance compared to one detector alone.
In the following, this acceptance is normed to 1 where each detector has its highest acceptance.
It is calculated at one instant in time and summed up for all \plenum-1 detectors, thus it is the instantaneous
signal acceptance relative to the highest acceptance of a single detector.

\subsection{Transient sources}
\label{sec:transients}
For the full source significance estimates, we employ a similar method as already described in~\cite{schumacher_plenum_2022}, where we investigated the capacity of \plenum to characterize the diffuse astrophysical neutrino flux.
Instead of looking at the full sky for the signature of the diffuse neutrino flux, we focus here on the point-like emission of single and multiple astrophysical sources. 
While this point-source method will be described and evaluated in more detail in an upcoming publication,
we investigate here specifically transient sources with various emission time windows on the scale of days to years.
The key observables to identify transient emission of neutrinos from astrophysical sources are the reconstructed energy, the arrival direction and the angular uncertainty, $\Psi=|\Vec{\Omega}_{\rm true} - \Vec{\Omega}_{\rm reco}|$.
We model the transient emissions as box time windows with the width of the respective flare.
This means that we do not account for any p-value corrections that might arise when the time or width of the flare time is not known.
Since our study is fully based on simulated data, we do not actually use the arrival time of neutrinos, but instead scale the expected number of signal and background events according to the width of the time window.
As long as the time windows are longer than a day (and multiples of full days), we do not need to account for the different field of views of the detectors in the Northern Hemisphere that rotate with respect to equatorial coordinates during one day;
Instead, we use the daily-averaged field of view for all detectors here.

We use the Poisson probability as the probability density function per bin, thus the generic formula for the full likelihood function is
\begin{equation}
\label{eq:diff_likelihood}
    \mathcal{L}({\rm data}~k~ |~{\rm model}~\mu) 
    = \prod_{{\rm bin\,}ij}^{N_{\rm bins}} \frac{\mu_{ij}^{k_{ij}}}{k_{ij}!}\cdot
    \exp \left( -\mu_{ij} \right) ~\text{with}~\mu_{ij} = \mu_{ij}^{\rm atm}(N_B) + \mu_{ij}^{\rm astro} (N_S).
\end{equation}
The binning is two-dimensional in reconstructed energy and angular uncertainty, $\Psi$.
The source declination enters only into the absolute event expectation via the declination-dependent effective area.
The model expectations, $\mu_{ij}$ per bin $ij$, are composed of an atmospheric (background) and an astrophysical (signal) neutrino expectation.
The two free parameters for all tests are the overall normalization factors, $N_{B,S}$, of the signal and background expectation.
As in our earlier approaches, we use Asimov data
to represent $k_{ij}$ in \autoref{eq:diff_likelihood}.
We use a likelihood ratio test and apply Wilks' theorem
to calculate the discrimination power between signal and background hypotheses.

\section{Results}
\label{sec:result}

\subsection{Sky coverage of neutrino alerts}
\label{sec:res_alert}

\begin{wrapfigure}{L}{.5\textwidth}
\vspace{-2em}
\includegraphics[width=0.5\textwidth]{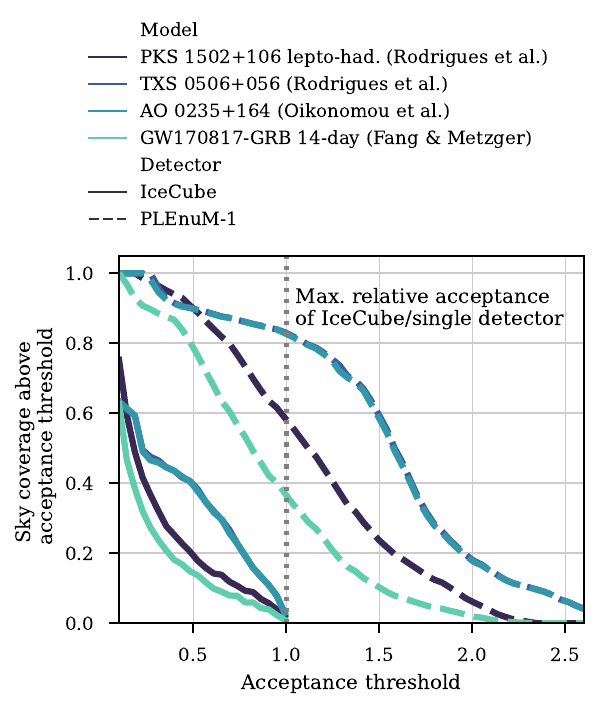}
\caption{Sky coverage above acceptance threshold for alert neutrinos with energy $E_{\rm reco} > \SI{100}{TeV}$ for four selected source models.
} 
\label{fig:alert-coverage}
\vspace{-0.5em}
\end{wrapfigure}
First, we estimate the capacity of \plenum-1 to detect high-energy neutrinos for real-time
correlation searches, as outlined in \autoref{sec:alerts}
\autoref{fig:inst_acceptance} shows the instantaneous and normalized acceptance of \plenum-1 for high-energy neutrinos, assuming the four selected source models as described in \autoref{sec:models}.
We see that the acceptance for highest-energy neutrinos produced by \pks and \gwgrb is more strongly peaked towards the horizons compared to \ao and \txs.
As a consequence, a smaller fraction of the sky can be covered with a high signal acceptance.
We quantify this sky coverage in a second step:
we use the signal acceptance of \autoref{fig:inst_acceptance} to calculate the sky coverage above a certain acceptance threshold,
which is shown in \autoref{fig:alert-coverage} for the four selected sources.
For example, IceCube can observe a source similar to \ao with at least 50\% of the maximum signal acceptance in 40\% of the sky, while \plenum-1 can cover 90\% of the sky at the same acceptance threshold.
Overall, the sky coverage with \plenum-1 improves by a factor of 2 to 5 above the 50\% acceptance threshold, depending on the assumed source spectrum.
Inversely, for a fixed sky coverage of 40\%, the acceptance threshold of \plenum-1 increases by a factor of 3.5 to 5 with respect to IceCube alone.
In addition, \autoref{fig:alert-coverage} shows the sky coverage based on the \pks and \gwgrb models falls more rapidly for higher acceptance thresholds compared to \ao and \txs.
This is again due to the signal acceptance being peaked more strongly towards the horizons.

\subsection{Discovery potential for selected sources}
\begin{figure}[h!]
    \centering
    \includegraphics[width=\textwidth]{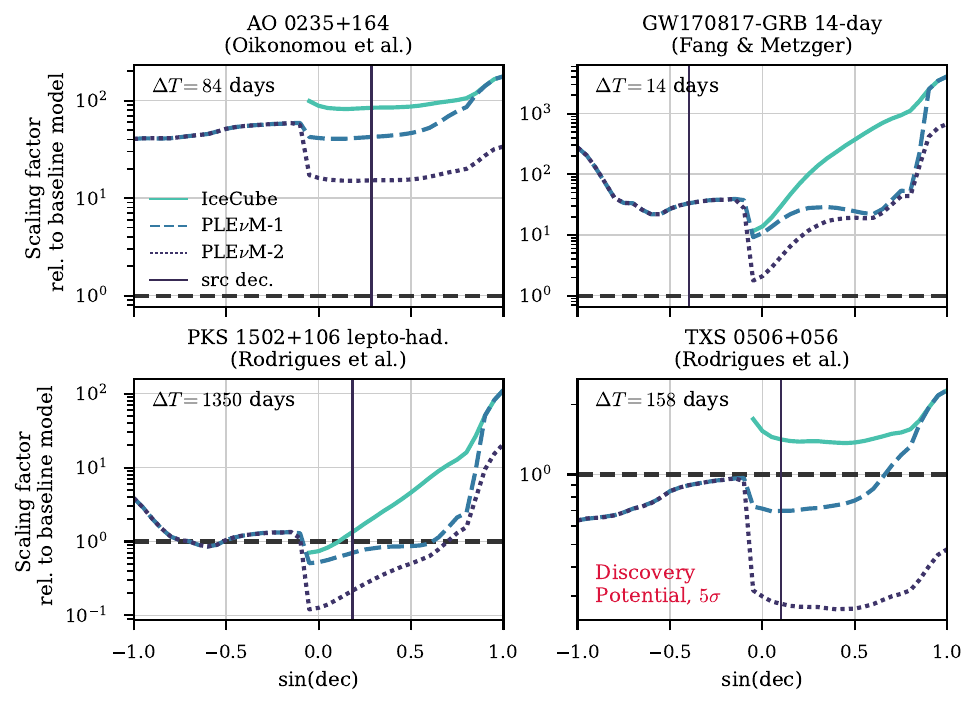}
    \caption{Discovery potential (DP) for selected sources as a function of sine of declination. The y-axis shows the factor by which the baseline models would need to be scaled to reach the DP.
    The vertical line marks the actual declination of the source.}
    \label{fig:model_DP_vs_dec}
\end{figure}
We calculate the discovery potential (DP) on the $5\sigma$ level for the four selected sources.
In addition, we use these sources as template models and calculate the DP at various declinations.
Note that we set the effective area of a single detector to zero above \SI{5}{deg} above the local horizon of all detectors,
as there the muon background becomes overwhelming and difficult to model using the public data.
\autoref{fig:model_DP_vs_dec} shows by what factor we would need to scale the template models of the four selected transient sources to obtain a $5\sigma$ discovery.
We find that with \plenum-2, a single flare similar to one of AO~0235+164 would need to be 10 to 50 times more luminous than predicted in the baseline model to reach $5\sigma$ significance.
The DP using the models of \gwgrb and \pks varies significantly depending on the source declination due to the strongly peaked acceptance close to the detectors' horizons.
This is a feature we already observed in the alert acceptances.
A source similar to \pks could be detected close to the horizon with IceCube, and almost over the whole sky with \plenum-1.
Also, a flare similar to the one detected in 2014-2015 of \txs can be observed over the whole sky with \plenum-1,
while \plenum-2 could improve the DP down to a fraction of the \txs flare flux.
Overall, we have a chance to discover faint sources only if these sources are observed over a long time span
or as a catalog of similar sources.

\subsection{Stacking of flaring FSRQs}
The flaring properties of the 145 blazars that are monitored with the Fermi-LAT have been investigated in~\cite{Yoshida:2022wac}. Using the Fermi-LAT flaring statistics of these sources and assuming that the neutrino flux scales in proportion to their $\gamma$-ray fluxes, here we study the detectability of neutrinos from the entire stacked sample during flaring epochs.

\begin{wrapfigure}{L}{.5\textwidth}
\includegraphics[width=0.5\textwidth]{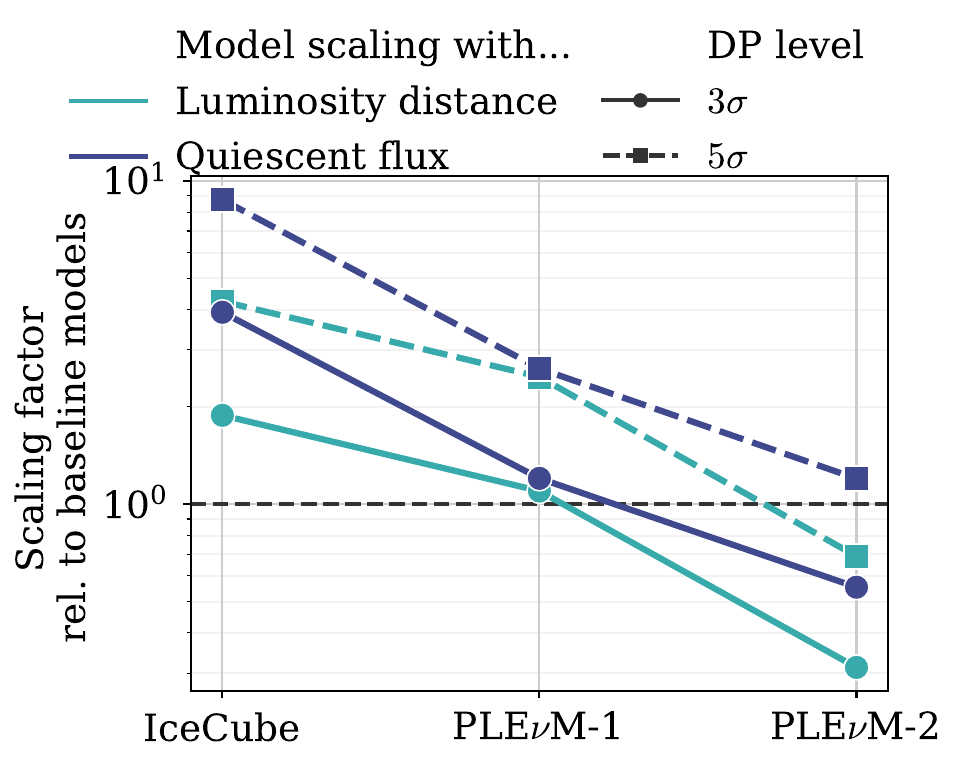}
\caption{Discovery potential on the level of $3\sigma$ and $5\sigma$ for the FSRQ stacking using two different model scaling factors.}
\label{fig:blazar_stacking}
\end{wrapfigure}
We select 106 flat-spectrum radio quasars (FSRQs) of this sample and add \txs as a  masquerading FSRQ~\cite{Padovani:2019_masquerading}.
As a neutrino flux expectation, we use the LMBB2b model for \txs in~\cite{keivani_multimessenger_2018} and scale its normalization for the other FSRQs.
We consider here two scaling factors per individual FSRQs:
one proportional to the inverse square of the luminosity distance and one proportional to the gamma-ray flux in the quiescent state.
We calculate the luminosity distance from the given redshift with the \texttt{Planck18} cosmology model of \texttt{astropy} ($H_0 = \SI{67.66}{km/s/Mpc}, \Omega_m = 0.3111, \Omega_\Lambda = 0.6889$)
~\cite{planck_collaboration_planck_2020,astropy2022}.
As the flare duration, we use the flare duty cycle as calculated in~\cite{Yoshida:2022wac} and scale it up to a total observation time of \SI{10}{years}. 
We have adopted the default threshold level of $6\sigma$ provided by~\cite{Yoshida:2022wac} to define a flare. 
At this significance level, \txs is assumed to be in a highly flaring state about 1\% of the time, which adds up to around \SI{37}{days} within \SI{10}{years} of observation time. 
In this sense, our results are conservative by a factor of a few, since selecting a lower threshold would have resulted in a flare duty cycle of $\mathcal{O}$(100 days).
\autoref{fig:blazar_stacking} shows the discovery potential when the expected neutrino flux of all FSRQs is summed up.
We find that it would take IceCube alone a minimum of thirty additional years of full operation to reach a $5\sigma$ signal. On the other hand,
with 10 years of integrated observation time, we find that
it might be possible to identify these FSRQs as neutrino sources on the $3\sigma$ level with \plenum-1 and on the $5\sigma$ level with \plenum-2.

\section{Discussion}
We presented in this proceeding the capability of \plenum to detect various different transient sources and source populations.
We show that, relative to the acceptance for high-energy neutrino alerts of IceCube, \plenum-1 improves the acceptance by up to a factor of 5.
In addition, also the instantaneous sky coverage for alerts can be improved by up to a factor of 5. 
While this will increase our chance to observe a high-energy neutrino in coincidence with a flaring source,
the discovery on the $5\sigma$ level will remain challenging for realistic blazar and sGRB models.
Only the most optimistic models for single sources will be detectable with \plenum-1 and 2.
However, our chances to detect a population of flaring sources are promising.
We find that the chance to detect a population of the brightest FSRQs is high, if we assume an optimistic model of the \txs flare of 2017. 
In summary, the detection of high-energy neutrinos and their sources proved challenging with IceCube due to the very low event rate for single sources at the highest energies.
It will remain challenging even with \plenum-1 and 2, but our chances to pinpoint the origin of the highest-energy neutrinos will improve significantly.

\printbibliography

\end{document}